\documentclass[preprint]{aastex}

\usepackage{graphicx}

\newcommand{\be}{\begin{equation}}
\newcommand{\ee}{\end{equation}}
\newcommand{\nn}{\mbox{} \nonumber \\ \mbox{} }
\newcommand{\ba}{\begin{eqnarray}}
\newcommand{\ea}{\end{eqnarray}}
\newcommand{\om}{\omega}
\newcommand{\Alfven}{ Alfv\'{e}n }

\newcommand{\B}{{\bf B}}

\newcommand{\RHESSI}{{\it RHESSI\,}}
\newcommand\etal{\textit{et al.\ }}
\newcommand\eg{\textit{e.g.\ }}

\newcommand\lo{\mathrel{\raise.3ex\hbox{$<$}\mkern-14mu\lower0.6ex\hbox{$\sim$}}}
\newcommand\go{\mathrel{\raise.3ex\hbox{$>$}\mkern-14mu\lower0.6ex\hbox{$\sim$}}}
\begin{document}
\date{}  
\title{ Polarization of prompt  GRB  emission: \\
evidence for electromagnetically dominated outflow}
\author{M. Lyutikov$^{1,2,}$\altaffilmark{3}, V.I. Pariev$^{4,5}$, and R.D. Blandford$^{6}$}
\affil{$^1$ Physics Department, McGill University, 3600 rue University,
Montreal, QC,\\Canada H3A 2T8 \\
$^2$ Canadian Institute for Theoretical Astrophysics,\\ 60 St. George, Toronto, Ont,  
M5S 3H8, Canada}
\altaffiltext{3}{lyutikov@physics.mcgill.ca}
\affil{$^4$ Department of Physics and Astronomy,
University of Rochester, Rochester, NY 14627 \\
$^5$ Lebedev Physical Institute, Leninsky Prospect 53,
Moscow 119991, Russia}
\affil{$^6$ Theoretical Astrophysics, California Institute of Technology,
Pasadena, California 91125}

\begin{abstract}
Observations by the {\RHESSI} satellite
 of large polarization of the prompt $\gamma$-ray emission 
from the Gamma Ray Burst GRB021206 \citep{coburn03}
imply 
 that the magnetic field coherence scale is  larger
than the size of the visible emitting region, $\sim R/\Gamma$, where
$R$ is the radius of the flow, $\Gamma$ is the associated Lorentz factor.
 Such  fields cannot be generated in a 
causally disconnected,  hydrodynamically dominated
outflow. Electromagnetic
models of GRBs \citep{lyutikov02}, 
in which  large scale, dynamically dominant, magnetic fields are  present in 
the outflow from the very beginning, provide a natural
explanation of this large reported linear  polarization.
We derive  Stokes parameters of  synchrotron emission
of a relativistically moving plasma with a given 
 magnetic field configuration
and calculate the  pulse averaged polarization fraction 
of the emission from a relativistically  expanding   shell carrying 
global toroidal
magnetic field.
For viewing angles larger than
$1/\Gamma$ the observed patch of the emitting shell  has almost homogeneous
magnetic field, producing a large fractional polarization
($56\% $ for a power-law energy 
distribution of relativistic particles 
$dn/d\epsilon \propto \epsilon^{-3}$). The  maximum
polarization  is  
smaller  than the theoretical  upper limit  for a 
stationary plasma in  uniform magnetic field
due to relativistic 
kinematic effects. 
\end{abstract} 

\section{ Introduction}

Origin of  magnetic fields  in Gamma Ray Bursts (GRBs) is  one 
of the central unresolved issues.  In the standard fireball scenario  
(\eg \citealt{piran99}, \citealt{meszaros02} and references therein), 
magnetic field does not play any  dynamical role. The near-equipartition field
invoked
in the emission region 
 is assumed to be generated locally  at relativistic shocks 
by plasma instabilities 
(\eg \citealt{medvedev99}). Initially,  the spatial scale of
 such fields is microscopically  small, 
of the order of the ion  skip depth, $\delta \sim c/\om_{p,i}$ 
($\om_{p,i}$ is the ion plasma  frequency). 
Though the typical scale of magnetic field
fluctuations may grow due to inverse cascade, even in the unlikely case
that such growth proceeds at the speed of light the resulting  polarization
is expected to be smaller 
 than $10 \%$ \citep[\eg][]{gruzinov99}. 

The 
recent detection by the {\RHESSI} satellite
of large polarization in the prompt $\gamma$-ray emission \citep{coburn03}
places severe constraints on the GRB models. 
It implies that magnetic field coherence scale is larger
than the size of the visible emitting region, $\sim R/\Gamma$, where 
$R$ is the distance form the center and  $\Gamma$
is a bulk Lorentz factor of relativistically expanding emission region.
Such  fields cannot be generated in a hydrodynamically dominated
outflow, which is causally disconnected on   large  scales. Thus, the 
 large scale  magnetic fields should 
be present in the outflow from the very beginning.
In fact, as we argue below, 
such fields must be dynamically dominant, carrying most
of the energy of the  outflow. 

Building  upon earlier models of electromagnetic explosions
 \citep[\eg][]{uso92,tho94,su96,mes97}, 
Lyutikov \& Blandford (2002, 2003) 
 developed an electromagnetic model  of GRBs
which 
assumes that  rotating, relativistic, stellar mass
progenitor (\eg ``millisecond magnetar'', \citealt{uso92})
 loses much of its spin energy in the  form of an
electromagnetically dominated outflow. A stellar  mass  relativistic
progenitor  is born
with angular velocity
 $\Omega \sim 10^4$ s$^{-1}$ and dynamo generated
 magnetic field of $ B_{\rm s} \sim 3 \times 10^{14}$ G.
Then the total rotational energy,
$E \sim I \Omega^2/2 \sim 5 \times 10^{52}$  erg (for a $1.4 \, M_\odot$
object) is available to power GRB bursts,
while the
 dipole spin-down luminosity
$L_P \simeq {B_{\rm s}^2 r_s^6\Omega^4/ c^3} \simeq  10^{49}\, { \rm ergs s^{-1}}$
is
about the
luminosities of cosmological $\gamma$-ray bursters.
In this model the energy to power the GRBs comes eventually from
 the rotational energy of the progenitor. It is first converted into
magnetic energy by the dynamo  action of the  unipolar inductor,
propagated in the form of Poynting flux dominated flow
and then dissipated at large distances from the sources.

A
rapidly spinning magnetar  with
a complicated field structure will form
a relativistic outflow. We suggest that magnetic field in the wind
 quickly rearrange
to become predominantly axisymmetric. There is a good precedent for this behavior in Ulysses observations
of the quiet solar wind \citep{ulys00} which
reveal that, despite the complexity
of the measured surface magnetic field, the field in the solar wind
quickly rearranges to form a good approximation to a \citet{par60} spiral.
The situation in the far field will then resemble that first analyzed
by \citet{gol69} and the characteristic scale length in the far field is the
cylindrical radius from a polar axis, rather than the wavelength.

During the relativistic expansion, most of the magnetic energy  
 carried by axisymmetric
 toroidal magnetic fields  is  concentrated in a thin shell with thickness
 $\Delta r \sim c t_s \sim 3 \times 10^{12}$ cm
inside a contact discontinuity separating  the
ejecta from  the shocked circumstellar material. At the  contact discontinuity 
 toroidal magnetic  field
balances the ram pressure of the circumstellar material
$ B_\phi \sim 4 \Gamma^2  \sqrt{\pi \rho_{\rm ext} c^2 } $. 
Both $B_\phi$ and $\Gamma$ depend on the  angle between a given point on the
shell and the polar axis, defined as the axis of rotation of the 
progenitor. This results in a non-spherical,
 relativistic expansion of the shell. 
In particular, for laterally balanced expansion $\Gamma \sim 1/\sin \theta$. 
The current carrying  shell becomes unstable due to development of
the current driven instabilities at a radius $\sim10^{16}$~cm. This leads to
acceleration of  pairs  which emit $\gamma$-rays by synchrotron radiation.

A distinctive feature of the
electromagnetic model is that the causal connection is better
than in the hydrodynamic models.
Initially, close to the central source
 the subsonic  flow is fully causally connected.
As the flow is  accelerated by magnetic (and partially by pressure) forces
it  becomes supersonic,  strongly relativistic and 
  causally disconnected
 over small
polar angles   $ \Delta \theta \sim 1/\Gamma$.
Later, magnetically dominated flows quickly reestablish  causal 
contact 
 over large polar angles and 
 become  {\it fully
causally-connected } again after a time $t_{\rm c} \sim t_s \Gamma^2 $,
where $t_s \sim 100$ sec is the source activity  life time. 
This is drastically different from hydrodynamic flows 
which remain causally disconnected over polar angles larger than $1/\Gamma$. 
Thus, 
during   expansion the  causal behavior of the flow resembles 
the behavior of cosmic fluctuations during inflation: 
as the flow expands,
 angular scales
$\sim\Gamma^{-1}$
 ``enter the horizon'' {\it re-establishing} causal contact that
was lost during acceleration.

To illustrate this behavior,
consider propagation of a  sound-type disturbance emitted by a point source
located on the  relativistically
moving shell at radius $R_{\rm em}$. 
Let the typical signal speed in 
the plasma rest frame be $\beta_s$.
In appendix \ref{causal} we show that for sub-Alfvenic ejecta
(magnetically dominated flows
can be strongly
relativistic, but still sub-Alfvenic!) a relativistically expanding 
shell  re-establishes a causal contact 
over the visible patch of $1/\Gamma$ in just  one dynamical time scale
 (after doubling in radius). 
If the  ratio
of the magnetic to particle energy density  in the cold,  magnetized plasma
 is
 $\sigma =  u_B/u_{\rm p} \gg 1 $
\citep{kennel84},  then the \Alfven (and fast magneto-sound) velocity 
is $\beta_{\rm A} = c \sqrt{ \sigma /(1+\sigma)}$. The
requirement that the expansion velocity be sub-Alfvenic
then implies that 
 $ c\Gamma \leq u_{\rm A} $ or  $\sigma \gg 1$.
Therefore,  the  condition that magnetic fields have a
 coherence scale larger than
$R/\Gamma$  requires that the 
magnetic fields  be energetically dominant in the flow.

Hydrodynamic  (\eg fireballs, \citealt{piran99}, or 
external shocks, \citealt{der02}, with $\sigma \ll 1$) 
 or hydromagnetic  models ($\sigma \sim 1$, \eg \citealt{sdd01}, \citealt{ds02},
\citealt{vlahakis03}) 
 could also  have a large scale ordered magnetic field (cannonballs,  
 \eg \citealt{dar03}, need to rely on energetically inefficient 
 Compton scattering and  strong flow inhomogeneities on the angular scale 
$\sim 1/\Gamma$,  \cite{coburn03}, section \ref{Disc}; 
we consider this prohibitively tight constraints).
 In all  cases,   the whole outflow
is in causal contact close
 to the source  and may have a large scale magnetic field 
which will be carried with the flow. 
 In hydro-dominated models, after the causal contact
is lost, different parts of the flow 
 cannot communicate and thus will evolve differently,
depending on the local conditions. Only under strict homogeneity of the
surrounding medium and of the ejecta the two causally disconnected  parts of the flow
will 
 have similar properties. On the other hand,
 since magnetically dominated outflows
can  quickly communicate information (\eg magnetic  pressure) 
 over  large polar angles, they 
can have  quasi-homogeneous  properties   despite
possible inhomogeneities in the circumstellar medium and in ejecta. 
 
Assumption of electromagnetically dominated 
flow must eventually break down, since
the ejecta need to dissipate magnetic energy to produce high energy emission.
In the emission region the 
plasma is expected  to be  close to equipartition as magnetic field is dissipated to
 accelerate electrons (generally, equipartition is needed for effective
emission). But unlike the hydrodynamic models
 where equipartition is reached by amplifying weak magnetic fields
at the shocks, in the  electromagnetically dominated model the
equipartition   is reached by dissipation of initially dominant
magnetic field, as, for example, happens in   solar flares.

In this paper we calculate the  Stokes parameters for the prompt  
GRB emission emerging in electromagnetic model
 as  a function of the viewing angle (angle between the line of sight
and the polar axis of the flow).
We assume that
magnetic field in the emission region  is dominated by the toroidal field and 
is concentrated  in a thin shell $\Delta R$ near the surface of the shell
expanding with the Lorentz factor $\Gamma(\theta)$. Synchrotron
 emission is produced by an isotropic
population of relativistic electrons with the power law
distribution in energy. In the present
work we calculate Stokes parameters averaged 
over the duration of the GRB pulse, deferring the time-dependent calculations
to a later work. One expects that the polarization fraction will be maximal
in the beginning of the pulse, 
slightly decreasing towards the end as larger emitting volumes
become visible. We set the speed of light to unity, $c=1$, in all the 
expressions to follow.

\section{Calculation of Stokes parameters}

\begin{figure}
\includegraphics[width=0.9\linewidth]{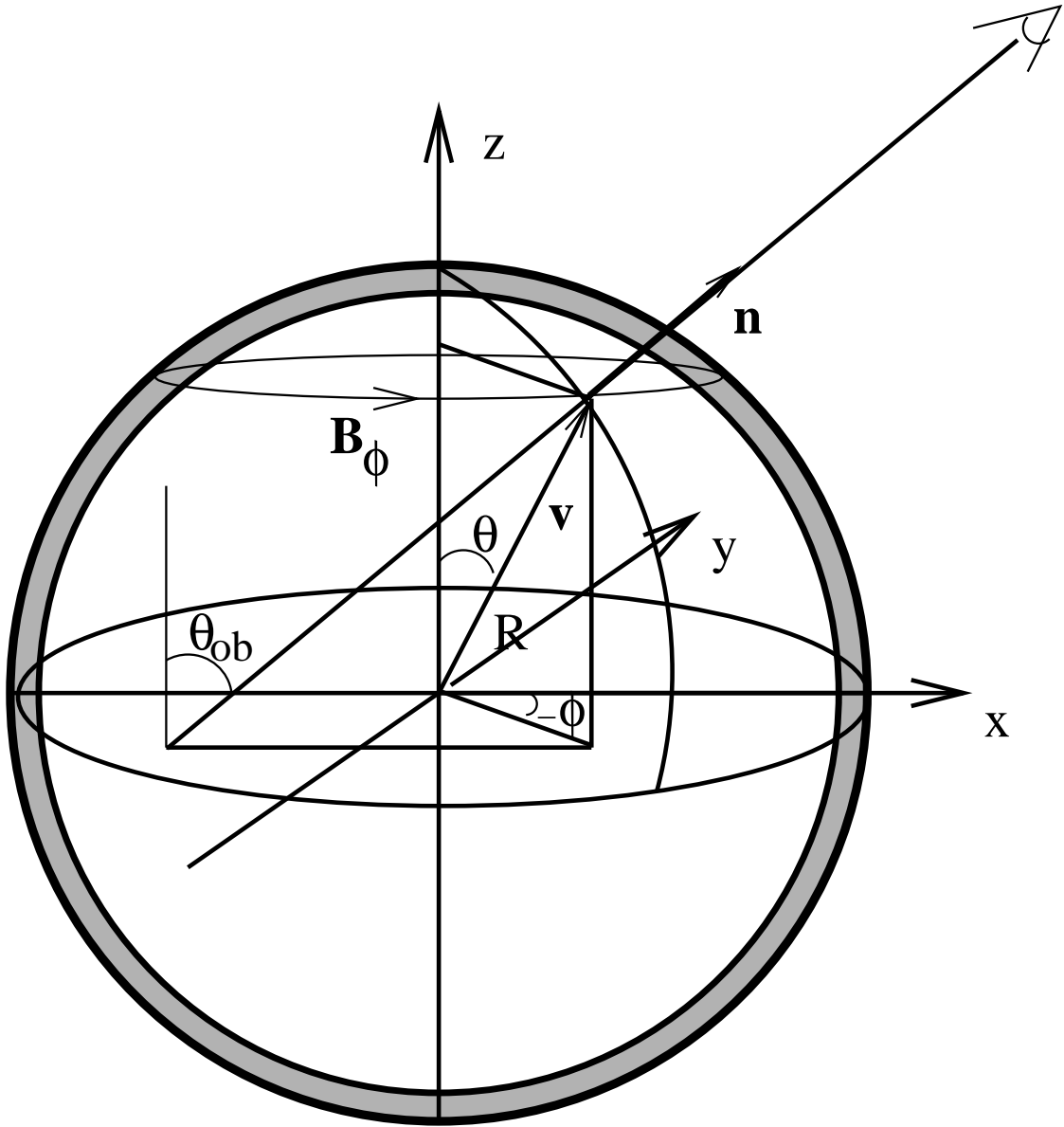}
\caption{
Geometry of the model.
 A narrow  shell $\Delta r  \sim t_s c \sim 3 \times 10^{12}$~cm,
 dominated by the toroidal magnetic field, expands quasi-spherically
with angle dependent velocity ${\bf v}(\theta)$.
 The 
observer is located at an angle $\theta_{\rm ob}$ 
with respect to the polar axis.
}
\label{Geometry}
\end{figure}

Consider a quasi-spherical thin emitting shell (Fig.~\ref{Geometry}) 
viewed by an observer.
Below we denote all quantities measured in the local frame comoving with
an emitting elementary volume with primes,
while unprimed notations refer to the quantities measured in the 
explosion frame.
Let $r$, $\theta$, and $\phi$ be the
spherical coordinates in the coordinate system centered at the center
of the shell, and $x$, $y$, and $z$ be the rectangular coordinates with 
the origin at the center of the shell. The symmetry axis of the shell 
is the $z$-axis. The toroidal magnetic field in the shell 
is in the $\phi$-direction. The observer is located in the $x$--$z$ plane.
The components of all vectors written below are the components with 
respect to the rectangular coordinate system $x$, $y$, and $z$.
The shell expands quasi-spherically
with an angle-dependent Lorentz factor $\Gamma(\theta)$.
An element of the shell moving radially
with the velocity $ {\bf v} = 
\beta \{ \sin \theta \cos \phi, \sin \theta \sin \phi, \cos \theta\}
$ emits a burst of synchrotron radiation  
in the direction of unit vector 
${\bf n} = \{\sin \theta_{\rm ob},0, \cos \theta_{\rm ob}\}$
when viewed in the observer frame.

Several key ingredients need to be taken into account 
(\eg  \citealt{cocke72}, \citealt{blandford79},
\citealt{bjornsson82}, \citealt{ginzburg89}).
First, the 
synchrotron emissivity depends on the direction between the emitted photon
and the magnetic field in the plasma rest frame. Second, as the 
emission is boosted by relativistic motion of the shell, the position angle 
of the linear polarization rotates in the ${\bf n} - {\bf v}$ 
 plane. 
\footnote[1]{This effect has been 
missed by all previous calculations of GRB polarization.}
The fractional polarization  emitted by each element remains the same, but
the direction of polarization vector of the radiation 
emitted by different elements within a visible shell
is rotated by different amounts. This leads to effective depolarization of the 
total emission. The theoretical maximum polarization fraction for homogeneous 
field  can be achieved  only for uniform plane-parallel velocity field.
 Third, integration along the line of sight (and over the 
emitting solid angle  for unresolved sources)
is better carried out in the laboratory frame, in order to take 
correct account of the  arriving times of photons.

We assume that the distribution function of emitting particles
in the frame comoving with an element of the shell 
is isotropic in momentum and is a power law in energy
\be
dn=K_{\rm e} \epsilon^{-p}d\epsilon dVd\Omega_{\bf p}\mbox{.}
\ee
Here $dn$ is the number of particles in the energy interval
$\epsilon,\epsilon+d\epsilon$, $dV$ is the elementary volume, 
$d\Omega_{\bf p}$ is the elementary solid angle in the direction of the
particle momentum ${\bf p}$, $K_{\rm e}=K_{\rm e}(r)$, $p={\rm constant}$.

In this paper we are interested in the polarization structure of the 
time integrated pulse of the emission, and not in its temporal properties. 
Hence, for simplicity
we approximate the $\gamma$-ray emissivity of the  shell as a
flash at some  time $t_0$ in the explosion frame, lasting 
for $\Delta t \ll \Delta R/c$, where $\Delta R \ll R_0$ is 
the thickness of the shell at the moment $t_0$ and $R_0$ is the 
radius of the shell at the moment $t_0$.
More complicated emission profiles may be easily accommodated.  In addition, we 
integrate over the observer time to get an average polarization of 
the pulse deferring time dependent calculations to a later work.

We also assume that the emission is optically thin and neglect possible 
plasma  propagation effects  (e.g.,
 depolarization of  radiation due to internal Faraday rotation
by low  energy electrons). Since the emitting particles are ultra-relativistic
and we neglect conversion of linear to circular polarization in plasma,
we do not have circular polarization in our model (Stokes $V=0$).
We also neglect a possible  tangled component
of the magnetic field  present in the emission region.
We assume that the emission 
originates in a geometrically thin
layer $\Delta R \ll R$ with the thickness $\Delta R$ independent on 
$\theta$ and neglect variation of the magnetic field and velocity across the  
layer. Given  these assumptions, our estimates provide an upper limit 
on the possible polarization.

Time-integrated Stokes parameters are calculated in appendix~\ref{J}
(Eqns. (\ref{Stokes})). Due to cylindrical symmetry of the model
the Stokes parameter ${\bar{U}}$ integrates to zero, so that the observed
averaged
polarization fraction is
\be
{{\Pi}} ={ |{\bar{Q}} | \over {\bar{I}}} = \frac{p+1}{p+7/3} 
{ \int \sin \theta d\theta d \phi
\,{\cal D}^{2+(p-1)/2} |B'\sin \chi'|^{(p+1)/2}
\cos 2{\tilde\chi} \over 
\int  \sin \theta d\theta d \phi
\,{\cal D}^{2+(p-1)/2} |B'\sin \chi'|^{(p+1)/2} }
\mbox{.}
\label{Pi}
\ee
Here 
$B'$ is the magnitude of the magnetic field ${\bf B}'$
in the frame of an element of the shell, 
$\chi'$ is the angle that the line of sight in the frame of an element 
of the shell, ${\bf n}'$, makes with the magnetic field ${\bf B}'$, 
and $\tilde{\chi}$ is the position angle of the electric field vector 
in the observer plane of the sky measured from some reference 
direction.  
Doppler boosting factor  is
$
{\cal D} = { 1/ \Gamma ( 1 - {\bf n} \cdot {\bf v})}\mbox{.}
$
For toroidal magnetic field  ${\bar{Q}} >0$, so that
the  observed polarization
vector is always along the projection of the flow axis on the plane 
of the sky.

Evaluation of  different quantities  in Eqn.~(\ref{Pi}) is an involved
exercise in Lorentz transformations. 
We assume that, 
in the 
shell frame, the  magnetic field is purely toroidal 
\be
\B'=  b_\phi(r,\theta) \hat{\B}'= 
b_\phi \{ -\sin \phi, \cos \phi, 0 \}\mbox{,}
\ee
where $ \hat{\B}'$ is the unit vector along $\B'$ in the 
radiating element frame 
 and 
$b_\phi$ is the magnitude of the field.
A photon propagating along the unit vector ${\bf n}$ in the 
explosion  frame is emitted along the direction with the unit 
vector ${\bf n}'$ in the radiating element frame:
\be
{\bf n}' =
\frac{ {\bf n} + \Gamma {\bf v} \left( {\Gamma\over \Gamma+1} ({\bf n} \cdot {\bf v}) -1\right)}{ \Gamma \left(1- ({\bf n} \cdot {\bf v})\right)}
\mbox{.}
\ee
Note, that 
${\bf n}', {\bf n}$ and $ {\bf v}$ lie in the same plane. 
The  
angle $\chi'$  between 
the  photon and  magnetic field in the radiating element frame is 
\be
\cos \chi' = {\hat{\B}' \cdot {\bf n}' } =
{ ( \hat{\B}' \cdot {\bf n}) +
 \Gamma  ( \hat{\B}' \cdot {\bf v})  \left( {\Gamma\over \Gamma+1} ({\bf n} \cdot {\bf v}) -1\right)
\over  \Gamma (1- ({\bf n} \cdot {\bf v})) }
\ee 
which gives
\begin{equation}
\sin^2\chi'=1-\frac{\sin^2\phi\sin\theta^2_{\rm ob}}{\Gamma^2(1-\beta\mu)^2}
\mbox{,}\label{sinchiprime}
\end{equation}
where $\mu=\cos\theta\cos\theta_{\rm ob}+\sin\theta\sin\theta_{\rm ob}\cos\phi$.

We also need to evaluate angle $\tilde{\chi}$ between a given direction in the 
observer 
plane and the polarization vector. This is not trivial since polarization vector
emitted by each element will experience  
rotation during Lorentz transformation
from the shell frame to the laboratory 
frame \citep{cocke72,blandford79}. Rotation of
the polarization vector is due to the rotation of the wave vector in the plane
containing vectors ${\bf n}$, ${\bf n}'$, and ${\bf v}$, and the requirement 
that the electric field of the wave
remains orthogonal to the  wave vector. 
Since   wave vectors of emitted waves experience
rotation by angles of the order of unity, this effect would lead to effective 
depolarization of emission from a  medium  with non-uniform velocity field
 even from a homogeneous magnetic field. 
In appendix \ref{etrnsform} we derive general relations for the Lorentz 
transformation of the polarization vector.

We choose to measure angle $\tilde{\chi}$ clockwise from the direction 
parallel to the projection of the axis of the flow 
on the plane of the sky. The unit vector in this direction is 
$\{-\cos\theta_{\rm ob}, 0, \sin\theta_{\rm ob} \}$.
We find (appendix \ref{etrnsform})
\ba &&
\cos \tilde{\chi} = { (1-\beta \mu) \cos \phi -
\beta  \sin \theta  \sin \theta_{\rm ob} \sin^2 \phi
\over
\sqrt{ (1-\beta \mu)^2 - \sin ^ 2  \theta_{\rm ob} \sin^2 \phi /\Gamma^2 } }
\mbox{,} \nn && 
\sin{\tilde\chi}= \frac{\sin\phi(\beta\cos\theta-\cos\theta_{\rm ob})}
{\sqrt{(1-\beta \mu)^2-\sin^2 \theta_{\rm ob} \sin^2 \phi /\Gamma^2  }}
\mbox{.}
\label{chi}
\ea 

In the ultra-relativistic limit, $\Gamma \gg 1$, general relations
simplify and it becomes possible to determine analytically the
maximum polarization fraction for  a given velocity field  
in the limit $\theta_{\rm ob} \Gamma \gg 1$
(appendix \ref{limits}). For $p=3$ we find 
$\Pi = 9/16$, in excellent agreement  with numerical calculations.

\begin{figure}
\includegraphics[width=.95\linewidth]{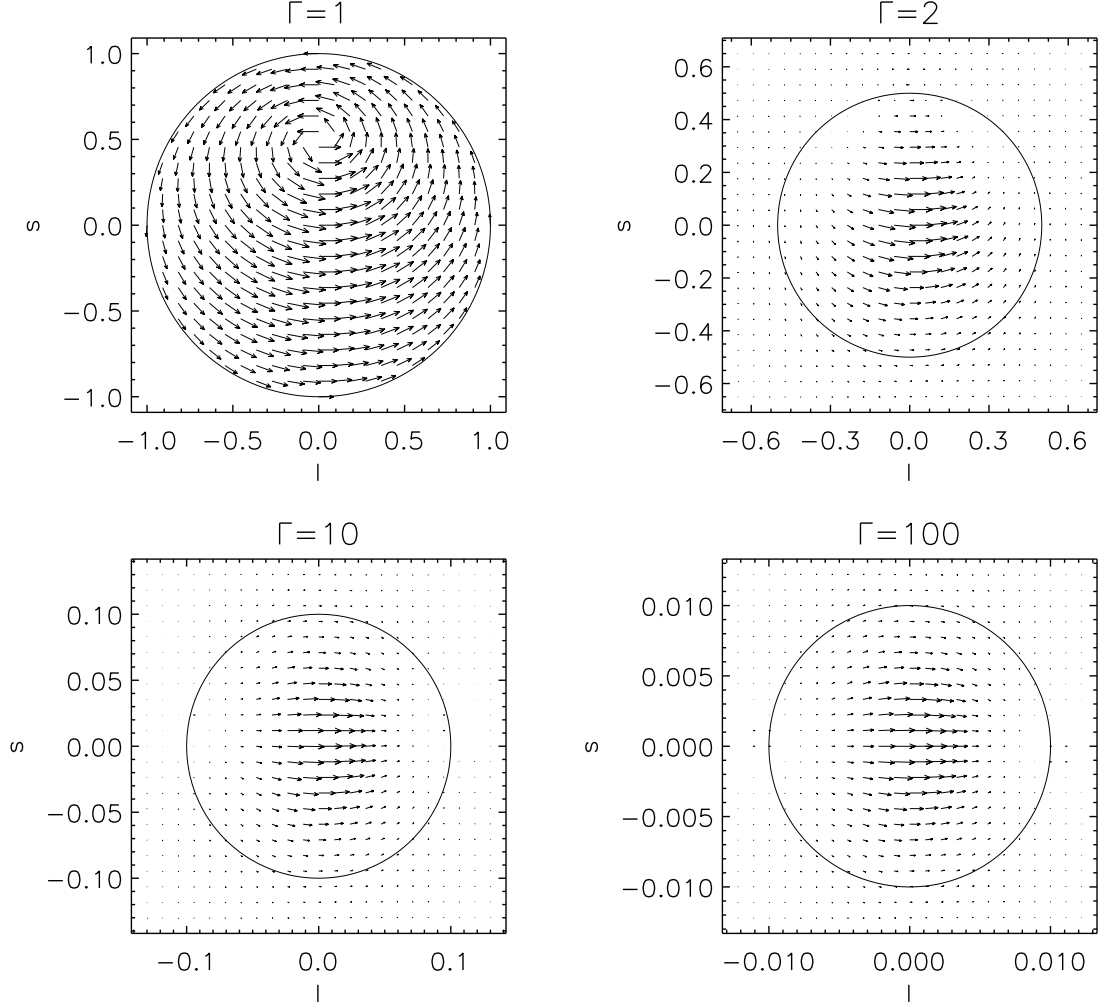}
\caption{Polarization map of the emission from GRB on the observer
plane of the sky in rectangular coordinates $l$ and $s$.
 Axis $s$ is directed parallel to the projection of the axis
of the shell on the sky. 
The observer line of sight makes $30^\circ$ angle with the axis
of the shell, $\theta_{\rm ob}=30^\circ$. Plots are made for four different
values of $\Gamma$ and $p=3$.  Solid circles  have radii
 $1/\Gamma$. 
As the intensity of radiation is
highly peaked in the area of the size $\sim 1/\Gamma$, we zoomed in
on this area in the plots with $\Gamma\gg 1$.}
\label{image}
\end{figure}

On Fig.~\ref{image}  we plot the map of the polarized
emissivity from the
shell moving with a constant $\Gamma$ and with constant $b_{\phi}$ as it
is seen by the observer on the plane of the sky.
 $l$ and $s$ are rectangular coordinates on the
plane of the sky centered at the projection of the center of the
shell. Axis $s$ is directed parallel to the projection of the axis
of the shell on the sky. $l$ and $s$ are normalized such that
the projection of the shell radius, $R_s$, is a circle of radius $1$
in the $l$--$s$ plane. Thus,
\be
l=\frac{1}{R_s}{\bf l}\cdot {\bf r}=\sin\theta \cos\phi
\mbox{,}\,\,\,
 s=-\frac{1}{R_s}{\bf r}\cdot ({\bf l}\times {\bf n}) =
- \cos\theta_{\rm ob}\sin\theta \cos\phi + \sin\theta_{\rm ob}\cos\theta
\mbox{,}\nonumber
\ee
where ${\bf l}= \{0,1,0\}$ is the unit vector along $l$.
The arrows on the plots in Fig.~\ref{image}  are directed
perpendicular to the unit vector in the direction of the electric field
of the wave, ${\hat {\bf e}}$, so that in the non-relativistic
limit, $\Gamma\to 1$,  the arrows  are  aligned with the magnetic field
${\bf B}$. The length of the
arrows is proportional to the synchrotron emissivity from the unit volume,
i.e., to the expression under the integral for ${\bar I}$ in
Eq.~(\ref{Stokes}). Actual observed intensity is modified by the
geometric factor proportional to the path of the ray inside the volume
of the shell. For $R_s-r \gg \Delta R$, the geometric factor
is $1/\mu$. For $\Gamma\gg 1$, Doppler boosting leads to the small effective
emitting area of the shell: $l \leq 1/\Gamma$ and
$s \leq 1/\Gamma$. 
Relativistic swing of polarization vector is also clearly
visible in Fig.~\ref{image}. Each patch of the
shell emits radiation with the same polarization degree,
$\Pi_{\rm max}=(p+1)/(p+7/3)$. Due to summation over the areas of
the shell with different directions of $\tilde\chi$, resulting polarization
degree becomes smaller than $\Pi_{\rm max}$.

We are now in a position to estimate polarization fraction
(\ref{Pi})  integrating  
Stokes parameters (\ref{Stokes}) over an expanding 
relativistic shell. 
Results for ${ \Pi}$ are shown in Fig.~\ref{stockes}. 
The parameter ${\bar Q}$ is zero for $\theta_{\rm ob}=0$ and is 
small for $\theta_{\rm ob} < 1/\Gamma$, 
because the polar axis falls within the visible patch in this case.
The magnetic field changes its direction within the visible patch and 
the resulting polarization is reduced.
The degree of polarization reaches
a limiting  value of tens percent when
observation angle is larger than $1/\Gamma$. 

\section{Discussion}
\label{Disc}

Large scale ordered magnetic fields produced at the central source provide
a simple explanation of the
 recent observations of highly  polarized GRB prompt
emission  by the  {\RHESSI} satellite \citep{coburn03}. 
In order to retain the  coherence of the  magnetic field
 on scales larger than the visible patch, $ \sim R/\Gamma$, the ejecta must be
electromagnetically dominated. 
The electromagnetic  model  suggested by Lyutikov \& Blandford
(2002, 2003)   provides a solution to the puzzle of how to produce
large coherent magnetic fields and  how to launch a blast wave that extends over an angular scale
$>>\Gamma^{-1}$ and where the individual parts are out of causal contact.
In the electromagnetic model, the magnetic fields are present in the outflow
from the very beginning, and the energy is transferred to the blast wave
by a magnetic shell which is causally connected at the end of
the coasting phase.

To prove this point we first  found general relations for transformation
of polarization direction of synchrotron emission produced
by relativistically moving source with a given magnetic field structure 
and calculated Stokes parameters for the  time averaged
synchrotron emission for a particular case of  relativistically 
expanding shell containing  toroidal magnetic fields. 
We find that for observing angles satisfying  $ \theta_{\rm ob}  \geq
1/\Gamma$ a large polarization fraction
$\Pi \sim 60 \%$ may be observed (the actual spectrum was not measured  
for GRB021206).
The position  angle of the polarization is fixed by the projection of the 
progenitor axis 
on the plane of the sky and thus should not change during the burst.

Another potential source of polarization could be Compton scattering 
of unpolarized $\gamma$-rays. If unpolarized $\gamma$-rays are initially 
beamed into a small-angle jet and are scattered by surrounding gas,
then polarized scattered $\gamma$-rays would be distributed nearly 
isotropically. This will require a much higher energy in the initial
$\gamma$-ray jet than the energy necessary if the narrowly beamed jet
is observed directly \citep{coburn03}, thus putting a much tougher 
requirement for the total energy budget of GRB. Compton scattering 
by relativistically moving wide angle envelope could also occur. In this 
case kinematics of the scattering is similar to the kinematics of the 
synchrotron emission considered in the present work. Therefore, the 
energetic requirements are also similar to the synchrotron mechanism. 
However, Compton scattered photons do not have preferred direction of 
polarization, which is set by the large scale magnetic field in the 
synchrotron case. The net polarization of scattered photons from 
uniform spherical shell averages to zero. High polarization can be 
observed only if the shell parameters ($\Gamma$ or electron density) 
vary significantly on the angular scale $\sim 1/\Gamma$. We are coming
back to highly collimated flow. Therefore, we
conclude that Compton scattering cannot account for the high degree 
of polarization of $\gamma$-rays emerging from a wide angle expanding
flow. We note, that synchrotron mechanism results in the electric 
vector of polarized emission directed parallel to the axis of the flow,
while the scattered $\gamma$-rays would be polarized in the direction
perpendicular to the axis of flow.

Several natural correlations between GRB polarization and other
parameters follow from the model and  can be tested with
future observations.
First, polarization fraction should decrease from the beginning to end of the
pulse as larger areas of the emitting shell become visible to the observer.
Second,
the maximum amount of polarization is related to the spectrum of emitting 
particles, being higher for softer spectra.
 This  points to a possible correlation between the amount of 
polarization and hardness of the spectrum. 

Our treatment of the prompt emission may also be related to  polarization 
of afterglows. If the field from the 
 magnetic shell  may be mixed in with the shocked
circumstellar material 
(similar to the  so called flux transfer events at the day side
of Earth magnetosphere), then  a comparably 
large fractional polarization may be observed in afterglows as well. 
In addition, since the preferred direction of polarization is always
aligned with the flow axis,  {\it the position angle should 
not be changing through the afterglow}
(if polarization is observed both in prompt and afterglow emission 
the position angle should be the same). 
Also, polarization should not be related to the "jet break" moment. 
This is in a stark contrast with  the jet model, 
in which  polarization  is seen only near the "jet break" times and
the position angle is predicted
to experience a flip during the "jet break" \citep{sari99}. For the same reason 
large {\it average} polarization cannot be due to particular viewing
geometry, as suggested by \citet{wax03}. 
Current polarization data show that in virtually all cases position
angle remains constant (\citealt{covino03a,covino03b,barth03,ber03}, 
see though \citealt{rol03}), 
while the 
amount of polarization does not show any correlation  with 
the "jet break".
This is consistent with the presence of  
large scale ordered magnetic fields in the afterglows.
(A  model of \cite{rossi02} of structured jets also predicts 
constant position angle, but since no large scale magnetic field is assumed the
polarization features are still related to the jet break times).

In our calculations we have neglected a random component of the
magnetic field which must be present in the emission region. 
Lyutikov \& Blandford  (2002, 2003)
 suggested that  $\gamma$-ray emitting electrons are accelerated
by current instabilities, somewhat similar to solar flares.
 Development of current instabilities should be accompanied 
by dissipation of magnetic fields and destruction of
the magnetic flux. These will generally add a random component to the ordered
magnetic field and will lead to a decrease in polarization
\citep{korchakov62}. The corresponding calculations are in progress.

An alternative model of GRBs that can feasibly give large scale magnetic fields
in the prompt emission region
 is the plerion model (\citealt{kon02,inoue03}, see also \citealt{lyu02}), 
which initially
was suggested for afterglows, but may also be extended to include the
prompt emission by external shock wave \citep{der02}. 
In this case,  the large scale  equipartition magnetic fields are created
ahead of the expanding GRB ejecta  by the preceding explosion 
of the ``supranova''  \citep{supranova}. 
Still this type of models faces similar 
causality/efficiency 
problem: if the plerion plasma is only at equipartition, $\sigma \sim 1$, 
it may be expected
to be inhomogeneous on $R/\Gamma$ scale; if it is strongly magnetized,
 $\sigma \gg 1$, then 
the  shocks will be   only weakly dissipative.

Implications of the {\RHESSI}  results, that GRB flows are 
electromagnetically-driven, may provide an important clue to the dynamics
of other  astrophysical sources like pulsars, (micro)quasars and  AGNs.
It is quite plausible that all these sources
 produce ultra-relativistic magnetically dominated   outflows
with low baryon density \citep{bla02}.
The flow evolution in all these systems may proceed in a similar way.
Energy, transported primarily by magnetic fields,
is dissipated far away from the source due to development of 
current instabilities. Particles are  accelerated in localized current
sheets by DC electric fields and/or electromagnetic turbulence
producing bright knots (in AGNs) and a variety of bright
spots in pulsar jet,  best observed in the Crab. 

\begin{acknowledgements}
VP acknowledges support from DOE grant DE-FG02-00ER54600.
\end{acknowledgements}

 After the submission of the paper additional
 observational details of the burst GRB021206 became available.
The spectral index of the burst  is  $\alpha \sim 0.6$; corresponding
particle index is 
 $p=2.2$ (Hajdas, private communication). The  maximum polarization
in our model is then
 $\sim 45\%$. In addition, there were several theoretical developments.
 Higher polarization (up to 100\%) can be achieved
if the particle distribution is not isotropic (Lazatti, private communication).
\cite{gra03} have performed  polarization calculations similar to ours, 
taking into account a random  component of the magnetic
field. He reached a similar conclusion that
 it is substantially easier to produce the polarization observed in
GRB 021206 from an ordered magnetic field. 
A small discrepancy between the 
 results 
 is explained by the difference in the duration of
the emission: while we assumed that the shell emits during time $\Delta t
\ll \Delta r$, \cite{gra03}  assumed that the emission is more prolonged,
$\Delta t \gg \Delta r \Gamma^2$.

\appendix

\section{Causal structure of relativistic  magnetized outflows}
\label{causal}

In this appendix we consider the casual structure of the relativistically
expanding magnetized shell. 
We wish to answer the  question: "If a surface of  relativistically
expanding magnetized shell is perturbed at a given  radius $R_{\rm em}$  
and zero polar angle, which points {\it on the surface of the shell}
will be affected after time $t$".

Consider propagation of a  sound-type disturbance emitted by a  point source
located on the  surface of  relativistically
moving shell at radius $R_{\rm em}$. Let the  signal speed in the plasma rest
 frame be $\beta_s$. 
For simplicity we assume that the shell is moving with constant velocity.
If in the plasma frame a wave is emitted in the direction $\theta_{\rm em}$
with respect to the flow velocity, 
\footnote{Propagation of waves in a non-uniformly moving
 medium will generally lead to a
change of the wave direction; for qualitative
estimates we neglect here this effect. This is well  justified in the 
strongly magnetized  limit 
$\sigma \rightarrow \infty$ and/or for small angles $\theta \leq 1$.}
 then in the laboratory frame
the  components of the wave  velocity   along and normal to the
bulk velocity are
\be
\beta_{s, lab, \parallel} =
{ \beta + \beta_s \cos \theta_{\rm em} \over
1+ \beta \beta_s \cos \theta_{\rm em}}
,\,\,\,
\beta_{s, lab, \perp} =
{  \beta_s \sin \theta_{\rm em} \over
\Gamma (1+ \beta \beta_s \cos \theta_{\rm em})}
\mbox{.}
\ee
The condition that a waves catches  the surface of the shell becomes
\ba && 
(R_{\rm em} + t \beta) \sin \theta = {  t \beta_s \sin \theta_{\rm em} \over
\Gamma (1+ \beta \beta_s \cos \theta_{\rm em})} \mbox{,}
\nn &&
(R_{\rm em} + t \beta) \cos \theta = R_{\rm em} + { \beta+ \beta_s \cos \theta_{\rm em} \over
1+ \beta \beta_s \cos \theta_{\rm em}} t \mbox{.}
\ea
Eliminating $\theta_{\rm em}$ we find
\be
2 (\beta \, \Gamma)^2  (1 - \beta_s^2) \sin ^2 {\theta \over 2} =
\sqrt{ 1 -{ \beta (2 R_{\rm em} +   \beta t ) \beta_s^2 
\over ( R_{\rm em} +   \beta t )^2 } }
+ { \beta \beta_s^2 \over  R_{\rm em} +   \beta t}  t  - 1 \mbox{.}
\label{thetacoh}
\ee
In case of isotropic relativistic fluid  with  internal sound  speed
 $ \beta_s =1/\sqrt{3}$,
Eq. (\ref{thetacoh}) implies that the maximum angle that sound waves can reach
as $t \rightarrow \infty$ is $ \theta \sim \sqrt{ \sqrt{6} - 2} /\gamma \sim 
0.67/\gamma $.
Thus, hydro-dominated relativistic plasma remains causally
disconnected   on scales $\theta \sim 1/\Gamma$ at all times.

In magnetically dominated medium the situation is drastically different.
Consider, for simplicity, cold magnetically dominated plasma.
If the  ratio
of the magnetic to particle energy density  in the flow is
 $\sigma =  u_B/u_{\rm p} \gg 1 $
\citep{kennel84},  the \Alfven  velocity is 
$\beta_{\rm A} =\sqrt{ \sigma /(1+\sigma)}$.
Eq. (\ref{thetacoh}) then becomes 
\be
2 (\beta \, \Gamma)^2 \sin ^2 {\theta \over 2} =
\sqrt{ (1+\sigma) \left( 1 + { R_{\rm em}^2 \sigma \over ( R_{\rm em} +   \beta t)^2} 
\right)} - \left(1 + { R_{\rm em} \sigma \over  R_{\rm em} +   \beta t} \right)
\mbox{.}
\label{1}
\ee
This implies that two points on the surface of the shell
  separated by an angle $\theta \ll  1 $  come into  causal contact 
after time
\be 
{c  t  \over R_{\rm em}} \sim \Gamma \theta  \sqrt{ 1+\sigma \over \sigma}
\mbox{.}
\ee
Thus,  in strongly magnetized medium, $\sigma \gg 1$, 
the visible patch of the shell with $ \theta \sim 1/ \Gamma$ re-establishes 
causal contact  in one dynamical time $t \sim  R_{\rm em} /c$. 

We may also invert Eq. (\ref{1}) to find time needed to establish
casual contact over angle $\theta$:
\ba &&
t= { 2 R_{\rm em} \beta \, \Gamma \sin {\theta \over 2} \over 
\beta \left( \sigma - 4 (\beta \, \Gamma)^2 \sin ^2 {\theta \over 2} ( 1+ (\beta \, \Gamma)^2 \sin ^2 {\theta \over 2})  \right) } \mbox{,}
\nn &&
\times
\left( \sqrt{ \sigma ( 1+ \sigma) (  1+ (\beta \, \Gamma)^2 \sin ^2 {\theta \over 2}) }+
\beta \, \Gamma \sin {\theta \over 2} \left( \beta \, \Gamma + 2 ( 1+ (\beta \, \Gamma)^2 \sin ^2 {\theta \over 2}) \right) 
\right)  \mbox{.}
\label{2}
\ea 
For $\sigma < 4 \Gamma^2( 1+ \Gamma^2)$ 
 the maximum causally connected region (for $t\rightarrow 
\infty$)  is finite
\be
\sin ^2 { \theta_\infty \over 2}   = {  \sqrt{  1+ \sigma} - 1 \over
2 (\beta \, \Gamma)^2 } \mbox{.}
\ee
For subsonic flow, $  \sqrt{ \sigma}  > \Gamma$, $\theta_{\infty}$
 becomes larger than $1/\Gamma$. 
For larger $\sigma$ whole shell  comes into a causal contact after time
\be
t= { 2R_{\rm em} \beta \, \Gamma \over 
\beta ( \sigma -  4 \Gamma^2( 1+ \Gamma^2))} 
\left( 2 (\beta \, \Gamma)^3 + \sqrt{ \sigma (1+ \sigma) (1+ (\beta \, \Gamma)^2)}
+ \beta \, \Gamma (2 + \sigma) \right)  \mbox{.}
\ee

It is also instructive to find 
 the emission angle $\theta_{\rm em} $ as a function of time
$t$ when an emitted wave catches with the surface of the shell:
\be
\cos \theta_{\rm em} = -
{1 \over  \beta  \beta_s (2 R_{\rm em} + \beta t) } 
\left( (R_{\rm em} + \beta t) -
 \sqrt{ ( R_{\rm em} + \beta t)^2 - t \beta (2 R_{\rm em} + \beta t) \beta_s^2} 
\right) \mbox{.}
\label{thetaem}
\ee

In an ultra-magnetized plasma (force-free plasma, $\sigma \rightarrow \infty$) 
relations simplify considerably. 
The region casually connected after time $t$
becomes
\ba &&
\sin { \theta  \over 2} = { 1\over 2  \Gamma} \sqrt{ t^2 \over 
R_{\rm em} (R_{\rm em} + \beta t) }  \mbox{,}
\nn &&
t = 2  R_{\rm em} \Gamma \sin { \theta  \over 2} 
\left( \beta \, \Gamma   \sin { \theta  \over 2} +
\sqrt{1+ (  \beta \, \Gamma )^2 \sin^2  { \theta  \over 2} } \right)  \mbox{.}
\ea
So that points separated by $1/\Gamma$ come into causal contact
after $ t = (1+\sqrt{5})  R_{\rm em} /2$ and  the whole
shell becomes causally connected after $t \sim 2  R_{\rm em} \Gamma^2$.
Emission angle 
(\ref{thetaem}) in the force-free case  then becomes
\be
\cos \theta_{\rm em} = - { t \over 2 R_{\rm em} + t \beta}  \mbox{.}
\ee
Note that in the plasma rest frame the waves which propagate  furthermost
 in polar
angles are emitted ``backward''
 {\it in the explosion frame}. 
It is  due to this fact  why they can ``beat" the commonly called
result that lateral velocity in the laboratory frame cannot be larger
than $c/\Gamma$. The latter is true only for waves propagating along the surface
of the shell (normal to the flow in the shell frame). We can understand then 
why hydrodynamic  sound waves cannot reach large polar angles: when emitted
``backward'',  they
are advected with the supersonically moving  flow.
 On the other hand, in subsonic  strongly magnetized plasma
fast magnetosound waves can outrun the flow and reach large polar
angles in the laboratory frame.

Thus, 
 if the  effective
signal velocity
(\Alfven velocity) in the bulk of the flow 
is larger than the expansion velocity,
 strongly relativistic  magnetized outflows 
 quickly become casually re-connected
over the visible patch $\Delta \theta \sim 1/\Gamma$. 
 For strongly subsonic flows, $\sigma \gg  \Gamma^2$,
points on the shell  separated by  $1/\Gamma$ come into causal
contact  on a dynamical time scale $t \sim R_{\rm em}$ 
(in a relativistic  hydrodynamical flow this never  happens).
  Since this time is fairly
short, the global dynamics of the flow is not very important which vindicates
our assumption of constant expansion velocity.

\section{Pulse-integrated  Stokes parameters}
\label{J}

The Stokes parameters are components of the polarization
tensor 
$J_{ls}=\frac12\left(\begin{array}{cc}
                              I+Q & U\\
                              U & I-Q
                              \end{array}\right).$
Here $x_l$ are coordinates in the plane perpendicular to ${\bf n}$
and there is no circular polarization. 
The pulse-integrated intensity 
\begin{equation}
{\bar J}_{ls}(\nu)=\frac{1}{D^2}\int
\, dT
\int dV  j_{ls}\left({\bf n},\nu,{\bf r},
T+\frac{r\cos\Theta}{c}\right) \label{v_8}\mbox{.}
\end{equation}
where $j_{ls}$ is emissivity,  $T= t - r \cos \Theta/c $ is the observer  time and 
 integration is over 
 the whole emitting region
in the explosion frame. 

We approximate emissivity as an instant flash at the moment $t=t_0$
with the duration $\Delta t$, $\Delta t \ll \Delta R /c$. We also
assume that the whole shell emits uniformly during the flash.
Then, the emissivity can be expressed as 
\begin{equation}
j_{ls}({\bf n},\nu,{\bf r},t)=
j_{ls}({\bf n},\nu,t_0)\delta(t-t_0)\Delta t
\left[H(r-R_0)-H(r-R_0-\Delta R)\right]
\label{v_10}\mbox{,}
\end{equation}
where $\delta(x)$ is Dirac delta function and $H(x)$ is a 
step function, $H(x)=1$ if $x>0$, $H(x)=0$ if $x<0$.
We first integrate in $T$ keeping all other independent variables
($r$, $\theta$, $\phi$) fixed:
\begin{eqnarray}
&& {\bar J}_{ls}(\nu)=\frac{1}{D^2}
\int_{0}^{2\pi}\,d\phi \int_{0}^{\pi}\,\sin\theta\,d\theta
\int_{0}^{+\infty}\, r^2\, dr \,
j_{ls}\left({\bf n},\nu,{\bf r},t_0\right)
\times \nonumber \\
&& \Delta t \left[H(r-R_0)-H(r-R_0-\Delta R)\right]
\label{v_9a}\mbox{.}
\end{eqnarray} 
Taking into account that $\Delta R \ll R$ and integrating 
in Eq.~(\ref{v_9a}) over $dr$, we obtain
\begin{equation}
{\bar J}_{ls}(\nu)=\Delta t \Delta R \frac{R_0^2}{D^2}
\int_{0}^{2\pi}\,d\phi \int_{0}^{\pi}\,\sin\theta\,d\theta
\,j_{ls}({\bf n},\nu,R_0,t_0) \label{v_11} \mbox{.}
\end{equation}
Lorentz transformation of the emissivity to the comoving frame 
with the element of the shell is
\begin{equation}
j_{ls}({\bf n},\nu,t_0)={\cal D}^2({\bf n}',t_0) 
j'_{l's'}({\bf n}',{\cal D}^{-1}\nu) \label{v_12}
\end{equation}
so we obtain
\begin{equation}
{\bar J}_{ls}(\nu)=\Delta t \Delta R \frac{R_0^2}{D^2}
\int_{0}^{2\pi}\,d\phi \int_{0}^{\pi}\,\sin\theta\,d\theta
\,{\cal D}^{2+(p-1)/2}  
j'_{l's'}({\bf n}',\nu) \label{v_13} \mbox{.}
\end{equation}
Using synchrotron expressions for $j'_{l's'}({\bf n}',\nu)$
in the comoving frame (e.g., \citealt{ginzburg89}) 
we obtain 
\ba &&
{\bar I}=\frac{p+7/3}{p+1}\kappa(\nu)\Delta R \Delta t 
\frac{R_0^2}{D^2 (1+z)^{2+(p-1)/2}}
\int_{0}^{2\pi}\,d\phi \int_{0}^{\pi}\,\sin\theta\,d\theta
\,{\cal D}^{2+(p-1)/2} |B'\sin \chi'|^{(p+1)/2} \mbox{,}
\nn &&
{\bar Q}=\kappa(\nu)\Delta R \Delta t 
\frac{R_0^2}{D^2 (1+z)^{2+(p-1)/2}}
\int_{0}^{2\pi}\,d\phi \int_{0}^{\pi}\,\sin\theta\,d\theta
\,{\cal D}^{2+(p-1)/2} |B'\sin \chi'|^{(p+1)/2}
\cos 2{\tilde\chi} \mbox{,}
\nn &&
 {\bar U}=\kappa(\nu)\Delta R \Delta t 
\frac{R_0^2}{D^2 (1+z)^{2+(p-1)/2}}
\int_{0}^{2\pi}\,d\phi \int_{0}^{\pi}\,\sin\theta\,d\theta
\,{\cal D}^{2+(p-1)/2} |B'\sin \chi'|^{(p+1)/2}
\sin 2{\tilde\chi} \mbox{,}
\nn &&
{\bar V}=0\mbox{,}
\label{Stokes}
\ea
where we  reinstated the cosmological factor $1+z$.
The function $\kappa(\nu)$ is
\be
\kappa(\nu) =
\frac{\sqrt3}4\Gamma_E\left(\frac{3p-1}{12}\right)
 \Gamma_E\left(\frac{3p+7}{12}\right)\frac{e^3}{m_{\rm e} c^2}\left[
   \frac{3e}{2\pi m_{\rm e}^3c^5}\right]^{(p-1)/2}\nu^{-(p-1)/2}K_{\rm e},
\ee
where
$e$ and $m_{\rm e}$ are the charge and mass of an electron, 
$\Gamma_E$ is the Euler gamma-function. 

The degree of polarization 
of the observed radiation pulse is expressed as 
$\Pi=\sqrt{{\bar Q}^2+{\bar U}^2}/{\bar I}$, giving Eq. (\ref{Pi}). 
The resultant position 
angle of the electric field $\tilde{\chi}_{\rm res}$ measured by the 
observer is found from
\be
 \cos 2\tilde{\chi}_{\rm res}=\frac{{\bar Q}}
{\sqrt{{\bar Q}^2+{\bar U}^2}}\mbox{,} \,\,\,
 \sin 2\tilde{\chi}_{\rm res}=\frac{{\bar U}}
{\sqrt{{\bar Q}^2+{\bar U}^2}}\mbox{,}
\quad 0\leq \tilde{\chi}_{\rm res} < \pi\mbox{.}
\ee
It can be checked that in our case under the change of $\phi$ to $2\pi-\phi$
in the integrals~(\ref{Stokes}) the value of ${\bar Q}$ is not changed,
and the sign of ${\bar U}$ is reversed. Therefore, 
the Stokes parameter ${\bar U}$ integrates out to zero. Consequently,
if ${\bar Q}>0$ then $\tilde{\chi}_{\rm res}=0$, if ${\bar Q}<0$ then 
$\tilde{\chi}_{\rm res}=\pi/2$. Thus, the observed electric vector can be 
either parallel or perpendicular to the projection of the axis of the 
flow on the plane of the sky. For a shell carrying only toroidal magnetic field
${\bar Q}>0$.

\section{Lorentz transformations of polarization vector}
\label{etrnsform}

In this appendix we first derive Lorentz transformations of polarization 
vector of the linearly polarized radiation emitted by
a relativistically moving plasma with a given magnetic field
and then find an angle  $\tilde{\chi}$ between a given direction 
(chosen later as a direction along the projection of the axis of the flow 
in the plane of the sky) and the direction of linear polarization of the waves 
for a spherically expanding shell. 

Let ${\bf n}'$ be a unit vector in the direction of a wave vector 
in the plasma rest frame, ${\hat {\bf B}}'$ be a unit vector along
the magnetic field in the plasma rest frame.
The electric field of a linearly polarized electromagnetic wave 
is directed along the unit vector ${\hat{\bf e}}' ={\bf n}' \times {\hat{\bf B}}'$  
and the magnetic field of the wave is along the unit vector 
${\hat{\bf b}}' = {\bf n}' \times  {\hat{\bf e}}'$, such that the Pointing flux along
${\hat{\bf e}}' \times {\hat{\bf b}}'$ is directed along ${\bf n}'$.
We will make a Lorentz boost to the explosion frame to find 
an electric field ${\bf e}$ there,  normalize it to unity, and 
project ${\bf e}$ on some given direction 
(\eg, along the projection of the flow axis on the plane of the sky).

Fields in the wave expressed in terms of the direction of a photon 
in the  explosion  frame  ${\bf n}$ are
\ba && 
{\hat{\bf e}}' = { {\bf n} \times {\hat{\bf B}}' \over 
\Gamma ( 1- {\bf n}\cdot {\bf v}) } +
{ 1+ \Gamma ( 1 - {\bf n}\cdot {\bf v}) \over 
(1+\Gamma)  ( 1- {\bf n}\cdot {\bf v}) } 
{\hat{\bf B}}' \times {\bf v}\mbox{,} 
\nn &&
{\hat{\bf b}}'=  - {\hat{\bf B}}' +
\left({ {\hat{\bf B}}'
 \cdot {\bf n} \over \Gamma^2 ( 1 - {\bf n}\cdot {\bf v})^2} -
{ 1+ \Gamma ( 1 - {\bf n}\cdot {\bf v}) \over
\Gamma (1+\Gamma) ( 1- {\bf n}\cdot {\bf v})^2 } {\hat{\bf B}}' \cdot {\bf v}
\right) {\bf n}
\nn &&
+
\left(
{ (1+ \Gamma ( 1 - {\bf n}\cdot {\bf v}))^2 \over
(1+\Gamma)^2  ( 1- {\bf n}\cdot {\bf v})^2 } {\hat{\bf B}}' \cdot {\bf v}
- { 1+ \Gamma ( 1 - {\bf n}\cdot {\bf v}) \over
\Gamma (1+\Gamma) ( 1- {\bf n}\cdot {\bf v})^2 } {\hat{\bf B}}' \cdot {\bf n}
\right) {\bf v} \mbox{.}
\ea
It may be verified that $ {\hat{\bf e}}' \times  {\hat{\bf b}}'$
is still directed along ${\bf n}'$.

Next, make a Lorentz transformation of $ {\hat{\bf e}}'$  back to the lab frame 
\be
{\bf e} = \Gamma \left( {\hat{\bf e}}' - {\Gamma \over \Gamma+1} 
({\hat{\bf e}}'\cdot {\bf v}) {\bf v} - {\bf v} \times {\hat{\bf b}}' \right) 
\ee
and normalize to unity. After some rearrangement we find
\ba &&
{\hat {\bf e}} ={  {\bf n} \times {\bf q}' \over 
\sqrt{ q^{\prime 2} - ( {\bf n} \cdot {\bf q}')^2} } \mbox{,}
\nn &&
{\bf q}' = {\hat{\bf B}}' +
 {\bf n}  \times (  {\bf v} \times  {\hat{\bf B}}') 
-{ \Gamma \over 1+\Gamma} ( {\hat{\bf B}}' \cdot {\bf v} ) {\bf v} \mbox{.}
\label{ee0}
\ea

\begin{figure}
\includegraphics[width=0.9\linewidth]{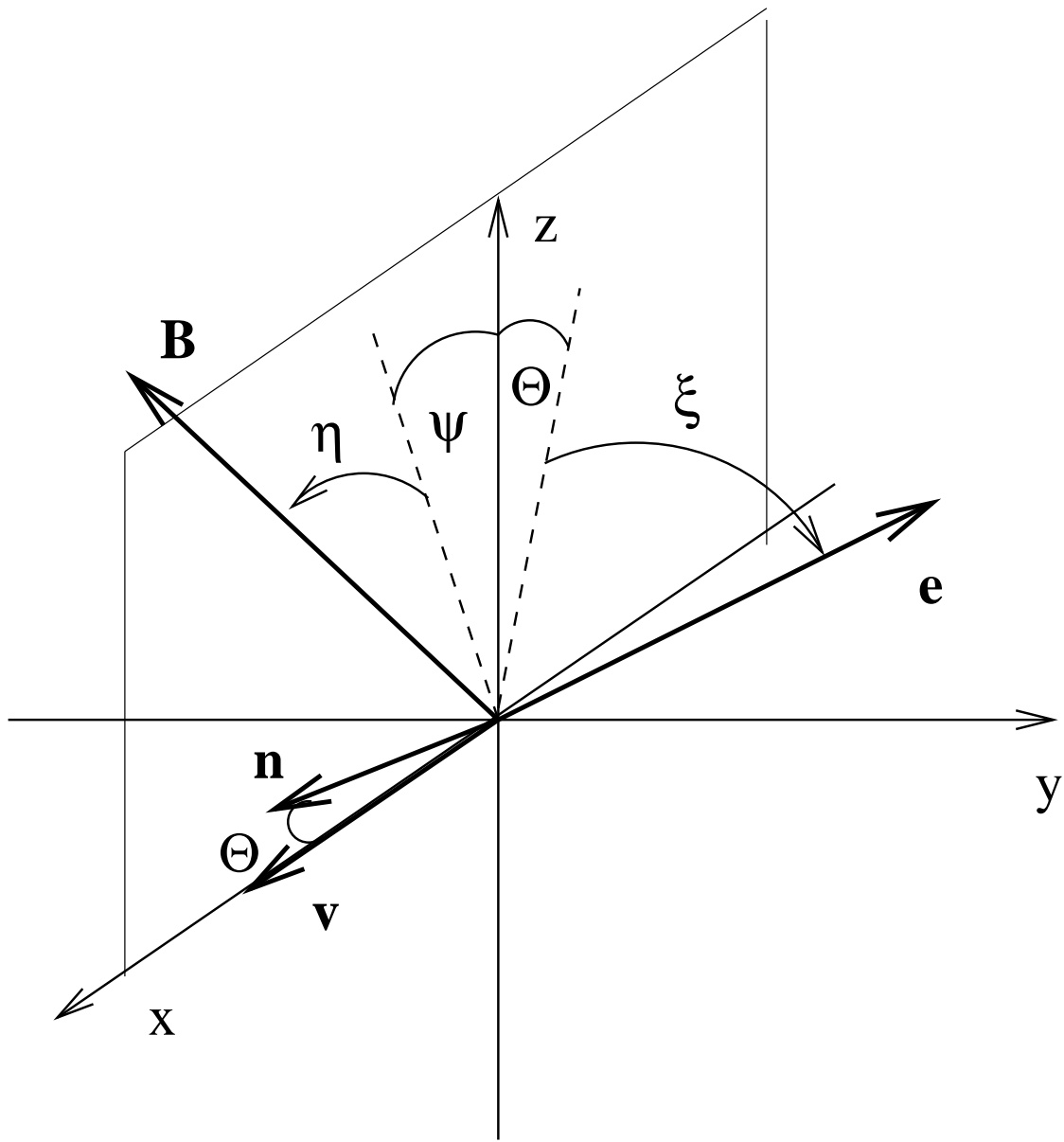}
\caption{ Characteristic swing of polarization angle due to relativistic
motion. In the frame 
aligned with ${\bf v}$, the electric field of the wave ${\bf e}$ 
and the observed magnetic field ${\bf  B}$  make
angles $\xi$ and $\eta$ with the plane containing ${\bf v}$ and ${\bf n}$, 
while their projections make 
angles $\Theta $ and $\psi$ with the axis $z$ perpendicular 
to ${\bf v}-{\bf n}$ plane (after \protect\citet{blandford79}).} 
\label{frame-bland}
\end{figure}

Finally, we may express the rest frame  unit vector 
${\hat{\bf B}}'$ in terms of the  laboratory frame unit vector 
${\hat{\bf B}}$. Assuming ideal MHD, there is no electric field in 
the rest frame of the plasma, ${\bf E}'=0$. Then, we obtain
\ba &&
{\hat{\bf B}} = 
{1 \over \sqrt{ 1 - ({\hat{\bf B}}' \cdot {\bf v})^2} } 
\left( {\hat{\bf B}}' - { \Gamma \over
 1+  \Gamma} ({\hat{\bf B}}' \cdot {\bf v}) {\bf v} \right)  \mbox{,}
\nn &&
{\hat{\bf B}}' = { (1+\Gamma) {\hat{\bf B}} + 
\Gamma^2 ({\hat{\bf B}} \cdot {\bf v}) {\bf v} \over
(1+\Gamma) \sqrt{ 1+  \Gamma^2 ({\hat{\bf B}} \cdot  {\bf v})^2 }}  \mbox{,}
\ea
 to get 
\ba &&
{\hat {\bf e}} ={  {\bf n} \times {\bf q} \over 
\sqrt{ q^2 - ( {\bf n} \cdot {\bf q})^2} } \mbox{,}
\nn &&
{\bf q} = {\hat{\bf B}} +
 {\bf n}  \times (  {\bf v} \times  {\hat{\bf B}})  \mbox{.}
\label{eee}
\ea
This is a general expression giving the 
 polarization vector in terms of the observed quantities
${\hat{\bf B}}$, ${\bf n}$ and ${\bf v}$.
If, for a moment,  we adopt 
  a frame aligned with the direction
of motion (Fig. \ref{frame-bland}), we find from (\ref{eee})
\be
\tan \xi = \cot \eta { \cos( \Theta+ \psi) - \beta  \cos \psi \over
1- \beta \cos \Theta}  \mbox{,}
\ee
reproducing Eq.~(16) in \citet{blandford79}.

In our particular case  
 ${\hat{\bf B}}' \cdot {\bf v}=0$, so that 
the fields in 
the rest frame of the emitting plasma element and the laboratory frame 
are aligned ${\hat{\bf B}}'={\hat{\bf B}}$.
The general relations then
simplify. Setting  ${\hat{\bf B}}' \cdot {\bf v}=0$
in Eq.~(\ref{ee0}) gives
\be
{\hat{\bf e}} =  { {\bf n} \times \left( {\hat{\bf B}} +
 \left({\bf n}  \times (  {\bf v} \times  {\hat{\bf B}})\right) \right) 
\over 
\sqrt{(1 - {\bf n}\cdot {\bf v})^2 - ({\hat{\bf B}} \cdot {\bf n})^2 /\Gamma^2}}
 \mbox{.}
\label{eee_simple}
\ee

Next we introduce a  unit vector ${\bf l}$
 normal to the plane containing  ${\bf n}$ and some given
direction (in our case the direction of the projection of the axis of the flow
to the plane of the sky). Then, 
\be
\cos \tilde{\chi} = {\hat{\bf e}}  \cdot ({\bf n}\times {\bf l}), \quad
\sin \tilde{\chi} = {\hat{\bf e}}  \cdot {\bf l} \mbox{.} 
\ee
Using Eq.~(\ref{eee_simple}) we find
\ba &&
\cos \tilde{\chi} = 
{  1 \over \sqrt{  1 - \left({ {\hat{\bf B}} \cdot {\bf n} \over 
\Gamma ( 1- {\bf n}\cdot {\bf v}) } \right)^2 } }
\left(
 {\hat{\bf B}} \cdot \left(  {\bf l}  + { ({\bf l}   \cdot {\bf v}) 
\over 1- {\bf n}\cdot {\bf v} } {\bf n} \right)  \right)  \mbox{,}
\nn &&
\sin  \tilde{\chi} =  {  \Gamma \over \sqrt{  1 - \left({ {\hat{\bf B}} \cdot {\bf n} \over
\Gamma ( 1- {\bf n}\cdot {\bf v}) } \right)^2 } }
\left(
{ ( {\hat{\bf B}} \cdot {\bf l} \times {\bf n}) 
} 
+
{ \Gamma ( {\bf n}\cdot {\bf v}) ( {\hat{\bf B}} \cdot  {\bf v} \times {\bf l})
\over (1+\Gamma)  } + 
{ \Gamma ( {\hat{\bf B}} \cdot  {\bf n}  \times  {\bf v} ) ({\bf l} \cdot  {\bf v}) \over 
(1+\Gamma)  } 
\right.
\nn &&
\left. + 
 { ( {\hat{\bf B}} \cdot {\bf n})   ({\bf l} \cdot  {\bf n}  \times  {\bf v} )
 \over  \Gamma  (1-  {\bf n}\cdot {\bf v})} \right) \mbox{.}
\ea

In our case
\ba && 
{\hat{\bf B}}=\{ -\sin \phi, \cos \phi,0 \},\,\,\,
{\bf n} =\{  \sin \theta_{\rm ob} , 0,\cos \theta_{\rm ob}  \},
\nn && 
{\bf v} = \beta 
 \{ \sin \theta \cos \phi, \sin \theta \sin \phi, \cos \theta \}
,\,\,\,
{\bf l}  = \{0,1,0\},
\nn &&
{\bf n}\cdot {\bf v} =  \beta  \mu
 ,\,\,\,
 \mu =  \cos  \Theta = 
 \cos \theta  \cos \theta_{\rm ob} + \sin \theta  \sin \theta_{\rm ob} \cos \phi
\mbox{,} 
\ea
which gives Eq.~(\ref{chi}).

\section{The ultra-relativistic limit.}
 \label{limits}

\begin{figure}
\includegraphics[width=0.9\linewidth]{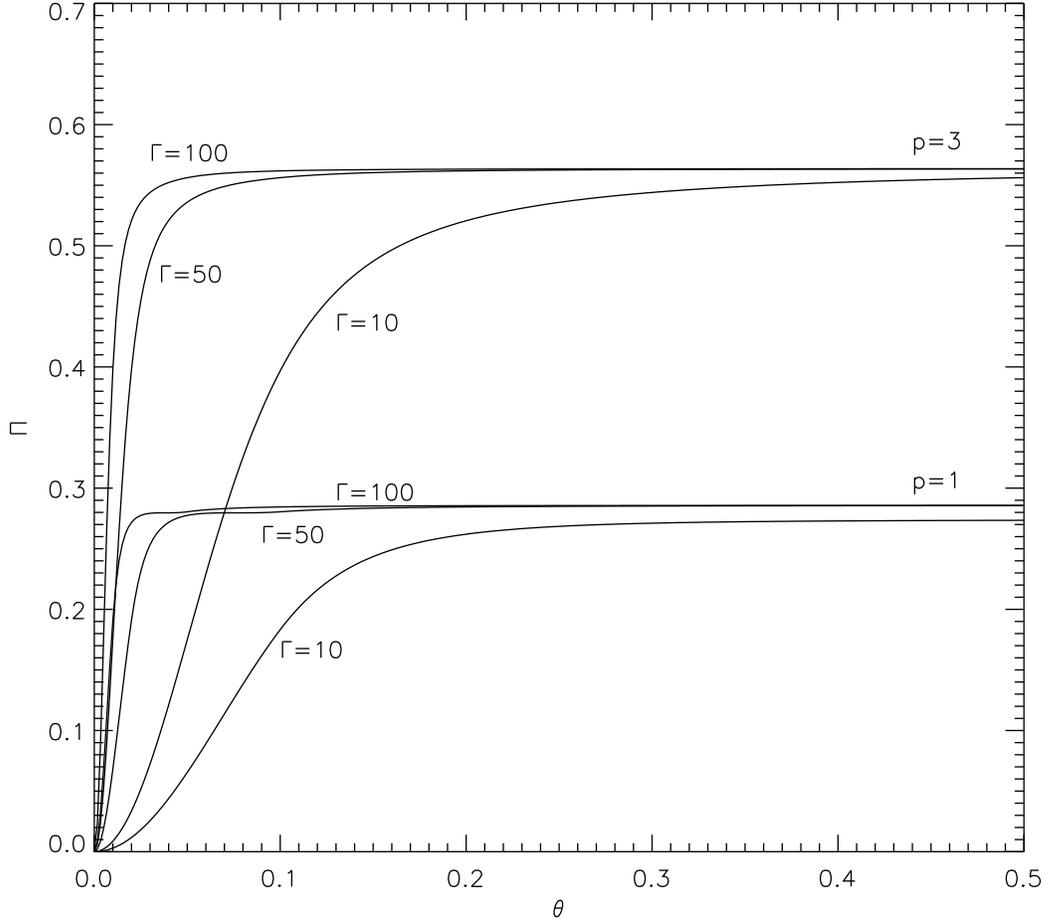}
\caption{Dependence of the polarization 
fraction $\Pi = Q/I$ on the viewing angle
$\theta_{\rm ob}$ for different Lorentz factors $\Gamma$ 
for isotropic expansion ($\Gamma(\theta)=
{\rm constant} =10, 50, 100$, right to left) 
and the power-law particle distribution 
$dn/d\epsilon = \epsilon^{-p}$; 
upper curves $p =3$, lower curves $p =1$ (for $p=2$ the asymptotic value 
is $43\%$). 
At $\theta_{\rm ob}=0$ polarization is zero, growing to large values when 
 $ \theta_{\rm ob} > 1/\Gamma$. Depolarization of emission due to
differential  rotation of the
 position angle
of the linear polarization  in the ${\bf n} - {\bf v}$ 
 plane reduces
the maximum possible polarization fraction below the 
theoretical limit for a homogeneous field.
}
\label{stockes}
\end{figure}

In the ultra-relativistic limit, when $\Gamma\gg 1$, the maximum 
polarization fraction for a given velocity field may be found analytically.
 Because of 
the Doppler boosting effect described by the factor ${\cal D}$, the 
contribution to the integrals in formula~(\ref{Pi}) comes from the 
small patch $\phi\sim 1/\Gamma$ and $\theta-\theta_{\rm ob}\sim 1/\Gamma$. 
We can introduce rescaled variables $\xi=\Gamma\phi$ and 
$\Psi=\Gamma(\theta-\theta_{\rm ob})$ and change the integration from $\theta$
and $\phi$ to $\xi$ and $\Psi$. The integration limits can be taken
from $-\infty$ to $+\infty$ for both $\xi$ and $\Psi$. 
Making expansions for $\Gamma\gg 1$,
$\xi\sim O(1)$ and $\Psi\sim O(1)$ the expression~(\ref{chi})
results in 
\begin{equation}
\cos 2\tilde{\chi}=\frac{(\Psi^2-\xi^2\sin^2\theta_{\rm ob}+1)^2-4\Psi^2
\xi^2\sin^2\theta_{\rm ob}}{(\Psi^2-\xi^2\sin^2\theta_{\rm ob}+1)^2 +
4\Psi^2 \xi^2 \sin^2\theta_{\rm ob}}
\label{chi17} \mbox{,}
\end{equation}
the expression for $\sin\chi'$ becomes
\begin{equation}
\sin^2\chi'=1-\frac{4\xi^2\sin^2\theta_{\rm ob}}{(1+\Psi^2+\xi^2\sin^2
\theta_{\rm ob})^2} \label{chipr17}\mbox{,}
\end{equation}
and the expression for ${\cal D}$ becomes
\begin{equation}
{\cal D}=\frac{2\Gamma}{\Psi^2+\xi^2\sin^2\theta_{\rm ob}+1}
\label{D17}\mbox{.}
\end{equation}
Note, that  
 the variable $\xi$ and $\sin\theta_{\rm ob}$ enter 
in the integrals~(\ref{Pi}) only in the combination $\xi\sin\theta_{\rm ob}$.
Therefore, by changing integration to a new variable 
$\xi_1=\xi\sin\theta_{\rm ob}$ the value of the integrals and $\Pi$ becomes
independent of $\theta_{\rm ob}$. Therefore, for $\Gamma\gg 1$ the polarization
degree is
 insensitive to $\theta_{\rm ob}$ as long as $\theta_{\rm ob} \gg 1/\Gamma$
(see Fig.~\ref{stockes}). 

\begin{figure}
\includegraphics[width=0.9\linewidth]{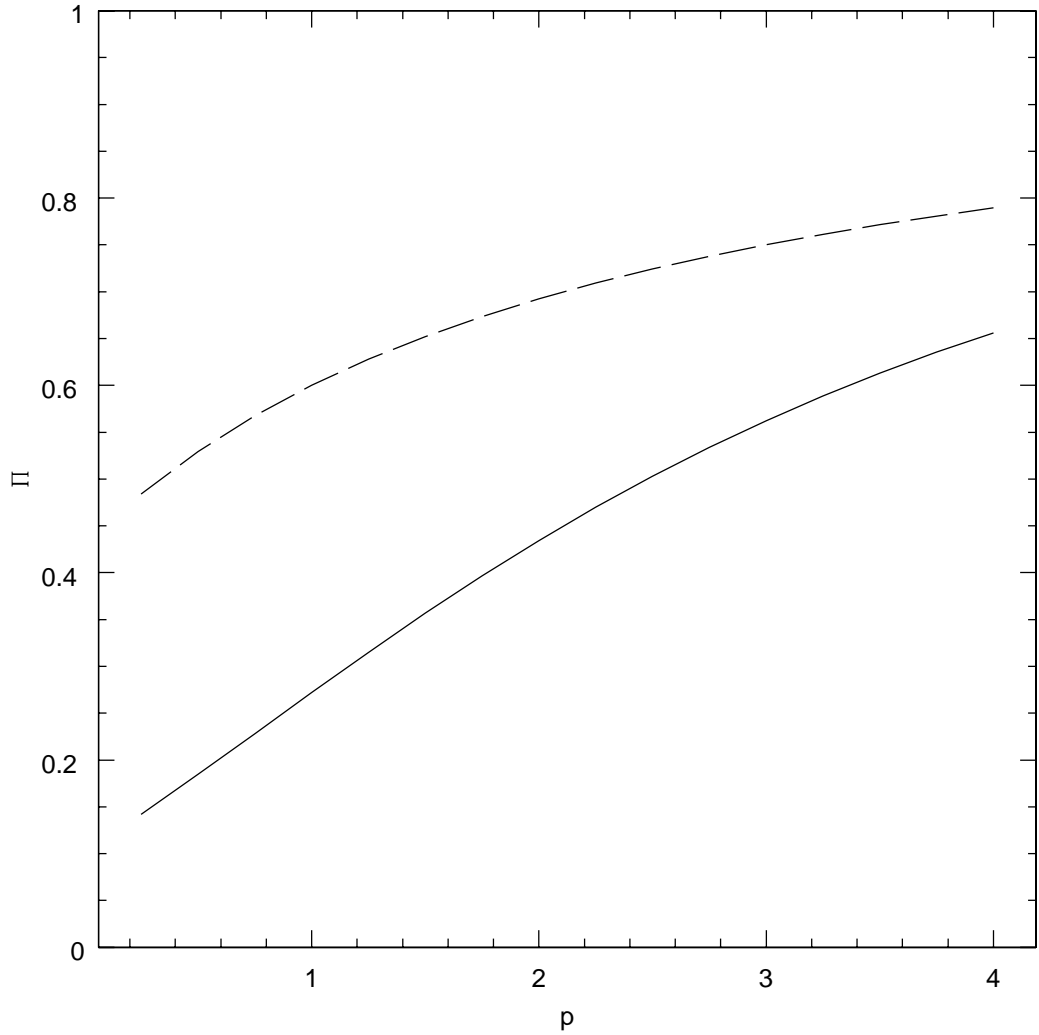}
\caption{Asymptotic value of polarization for $\Gamma \rightarrow \infty$ as a 
function of $p$ for spherically divergent flow (solid line), maximum polarization 
for a homogeneous stationary magnetic field $\Pi_{\rm max}$ (dashed line).}
\label{Pimax}
\end{figure}

Further, it is convenient to switch to the integration in ``polar'' 
coordinates $\sigma$ and $\tau$ in the plane $\xi_1$ and $\Psi$,
which are introduced 
according to $\sigma=\xi_1^2+\Psi^2$, $\xi_1=\sqrt{\sigma} \cos(\tau/2)$, and
$\Psi=\sqrt{\sigma} \sin(\tau/2)$. After some algebra we obtain
\begin{equation}
\Pi=\frac{p+1}{p+7/3}
\frac{\displaystyle\int_0^{\infty}\,d\sigma\,\int_0^{2\pi}\,d\tau 
(1+\sigma^2 \cos 2\tau - 2\sigma\cos\tau)
\frac{(1+\sigma^2-2\sigma\cos\tau)^{(p-3)/4}}{(1+\sigma)^{2+p}}
}
{\displaystyle\int_0^{\infty}\,d\sigma\,\int_0^{2\pi}\,d\tau 
\frac{(1+\sigma^2-2\sigma\cos\tau)^{(p+1)/4}}{(1+\sigma)^{2+p}}
}
\label{Pi17}\mbox{.}
\end{equation}
For $p=3$ expression~(\ref{Pi17}) gives $\Pi=9/16 \approx 56\%$. This 
value is the value of the horizontal asymptotic of $\Gamma(\theta_{\rm ob})$ 
curve for $p=3$ in Fig.~\ref{stockes}. For arbitrary  $p$ 
asymptotic values  of the polarization are plotted in Fig. \ref{Pimax}.

\begin{thebibliography}{}

\bibitem[Barth \etal(2003)]{barth03} 
Barth,~A.J. \etal 2003, \apj Lett., 584, 47

\bibitem[Bersier \etal(2003)]{ber03} Bersier,~D. \etal 2003,
\apj  Lett., 583, 63

\bibitem[Bjornsson(1982)]{bjornsson82} Bjornsson,~C.-I. 1982, \apj, 
260, 855 

\bibitem[Blandford(2002)]{bla02}Blandford, R. D. 2002, "Lighthouses
of the Universe", ed. R. Sunyaev,  Berlin:Springer-Verlag

\bibitem[Blandford \& K\"onigl(1979)]{blandford79} Blandford,~R.D., 
\& K\"onigl,~A. 1979, \apj, 232, 34

\bibitem[Coburn \& Boggs(2003)]{coburn03} Coburn,~W., \& Boggs,~S.E.
2003, \nat, 423, 415

\bibitem[Cocke \& Holm(1972)]{cocke72} Cocke,~W.J., \& Holm,~D.A. 
1972, Nature Phys. Sci., 240, 161

\bibitem[Covino et~al.(2003a)]{covino03a} Covino,~S., et~al. 
2003, \aap, 400, 9

\bibitem[Covino et~al.(2003b)]{covino03b} Covino,~S., Ghisellini,~G., 
Lazzati,~D., \& Malesani,~D. 2003, astro-ph/0301608

\bibitem[Dar(2003)]{dar03} Dar,~A., 2003, astro-ph/0301389

\bibitem[Dermer(2002)]{der02} Dermer,~C.D., 2002, \apj, 574, 65

\bibitem[Drenkhahn \& Spruit(2002)]{ds02}
 Drenkhahn, G.; Spruit, H. C., 2002, A\&A, 391, 1141

\bibitem[Ginzburg(1989)]{ginzburg89} Ginzburg~V.L. 1989, 
Applications of Electrodynamics in Theoretical Physics and
Astrophysics. (New York: Gordon and Breach Science Publishers)

\bibitem[Goldreich \& Julian(1969)]{gol69}Goldreich,~P. \& Julian,~W.H.,
1969, \apj, 157, 869

\bibitem[Granot(2003)]{gra03} Granot, J.,  submitted to ApJL 
(astro-ph/0306322)

\bibitem[Gruzinov \& Waxman(1999)]{gruzinov99} Gruzinov,~A., \& Waxman,~E. 
1999, \apj, 511, 852 

\bibitem[Inoue \etal(2003)]{inoue03} 
Inoue,~S., Guetta,~D., Pacini,~F. 2003, \apj,
583, 379

\bibitem[Kennel \& Coroniti(1984)]{kennel84} Kennel,~C.F., \& 
Coroniti,~F.V. 1984, \apj, 283, 694

\bibitem[K\"onigl \& Granot(2002)]{kon02} K\"onigl,~A. \& Granot,~J. 
2002, \apj, 574, 134

\bibitem[Korchakov \& Syrovatskii(1962)]{korchakov62} 
Korchakov,~A.A., \& Syrovatskii,~S.I. 1962, 
Soviet Astronomy, 5, 678 


\bibitem[Lyutikov(2002)]{lyu02}Lyutikov, M., 2002, Phys. Fluids, 14, 963

\bibitem[Lyutikov \& Blandford(2002)]{lyutikov02} Lyutikov,~M., \& 
Blandford~R.D. 2002, in "Beaming and Jets in Gamma Ray Bursts", 
 R. Ouyed, J. Hjorth and A. Nordlund, eds.,
 astro-ph/0210671

\bibitem[[Lyutikov \& Blandford(2003)]{lyutikov03} Lyutikov,~M., 
\& Blandford,~R.D. 2003, in preparation

\bibitem[McComas \etal(2000)]{ulys00}
McComas,~D.J. \etal, 2000, \jgr, 105, 10419

\bibitem[Medvedev \& Loeb(1999)]{medvedev99}
Medvedev,~M.V., \& Loeb,~A. 1999, \apj, 526, 697

\bibitem[M\'esz\'aros \& Rees(1997)]{mes97}M\'esz\'aros, P., Rees, M. J.
1997, \apj  Lett., 482, 29

\bibitem[M\'esz\'aros(2002)]{meszaros02}
M{\' e}sz{\' a}ros,~P. 2002, \araa, 40, 137

\bibitem[Parker(1960)]{par60}Parker,~E.N. 1960, \apj, 132, 821

\bibitem[Piran(1999)]{piran99} Piran,~T. 1999, Phys. Reports, 314, 575

\bibitem[Rossi \etal(2002)]{rossi02} Rossi,~E.,  Lazzati,~D.,
Salmonson,~J.D.,  Ghisellini,~G. 2002, astro-ph/0211020

\bibitem[Rol \etal(2003)]{rol03} Rol,~E. \etal 2003, atro-ph/0305227

\bibitem[Sari(1999)]{sari99} Sari,~R. 1999, \apj  Lett., 524, 43 

\bibitem[Smolsky \& Usov(1996)]{su96} Smolsky~M.V., Usov~V.V. 1996, ApJ, 461, 858

\bibitem[Spruit \etal(2000)]{sdd01} Spruit~H.C., Daigne~F., Drenkhahn~G.
2001, \aap, 369, 694

\bibitem[Thompson(1994)]{tho94}Thompson, A. C. 1994, MNRAS, 270, 480

\bibitem[Usov(1992)]{uso92} Usov,~V.V. 1992, Nature, 357, 472

\bibitem[Vietri \& Stella(1999)]{supranova} Vietri,~M., Stella,~L.,
1999, \apj  Lett., 527, 43

\bibitem[Vlahakis \& K\"onigl(2003)]{vlahakis03} 
Vlahakis,~N. and Konigl,~A. 2003,  astro-ph/0303482

\bibitem[Waxman(2003)]{wax03} Waxman,~E. 2003, Nature, 423, 388 
\end {thebibliography}
\end{document}